\documentclass[12pt,preprint]{aastex}

\begin{document}


\title{Gamma-ray emission from strongly magnetized pulsars}


\author{Anatoly E. Shabad}
\affil{P.N. Lebedev Physics Institute, Moscow 117924, Russia}

\and

\author{Vladimir V. Usov}
\affil{Center for Astrophysics, Weizmann Institute, Rehovot 76100,
Israel}




\begin{abstract}
In a strong magnetic field, $B\gtrsim 4\times 10^{12}$ G,
$\gamma$-rays emitted nearly along curved field lines adiabatically
convert into bound electron-positron pairs (positronium atoms)
rather that decaying into free pairs. This process may modify the
polar gaps of strong magnetized pulsars. Unlike free pairs, such
bound pairs do not screen the electric field component along the
magnetic field in the polar gaps. As a result the total power
carried away by both relativistic particles and radiation, from the
polar gap into the pulsar magnetosphere, may increase significantly
(up to a few tens times) in comparison with the conventional polar
gap models where creation of bound pairs is ignored, and it may be a
substantial fraction of the spin-down power. We demonstrate that the
total power of the modified polar gaps may be enough to explain the
observed non-thermal luminosities of all known strongly magnetized,
$\gamma$-ray pulsars.

\end{abstract}





\section{Introduction}

After the {\it Compton Gamma Ray Observatory (CGRO)} ended its
activity in 2000, seven rotation-powered $\gamma$-ray pulsars were
known (e.g., Thompson et al. 1997). More recently, the {\it AGILE}
telescope reported the detection of two other radio pulsars in
$\gamma$-rays. The {\it Fermi Gamma-ray Space Telescope} was
successfully launched in 2008. After two years of observations by
{\it Fermi}, the number of know $\gamma$-ray pulsars increased
dramatically, to more than sixty, and continue increase (Ray \&
Saz-Parkinson 2010). For known $\gamma$-ray pulsars the maximum of
the radiated power is concentrated in the $\gamma$-ray range, except
for the Crab pulsar and PSR B1509-58, the power of which peaks in
hard X-rays. The pulsar luminosities $L_\gamma$ in $\gamma$-rays are
a substantial fraction of the spin-down power, $\eta_\gamma =
L_\gamma/\dot E_{\rm rot}\sim 10^{-2}-1$ (Abdo et al. 2010), where
$\dot E_{\rm rot}=4\pi^2 I \dot P P^{-3}$ is the spin-down power,
$I$ is the moment of inertia of the neutron star (generally taken to
be $10^{45}\,{\rm g\,cm}^2$), $P$ its spin period, and $\dot P$ its
spin-down rate.

\par
The starting point common to all viable models of pulsars is that
strong electric fields are generated in the magnetospheres of
rotating neutron stars (e.g., Michel 1991). The electric field
component, $E_\parallel =({\bf E\cdot B})/|{\bf B}|$, along the
magnetic field, ${\bf B}$, is non-zero and may accelerate particles
to ultrarelativistic energies. The accelerated particles emit
$\gamma$-rays due to curvature emission and other processes. Some of
the $\gamma$-rays are converted into $e^+e^-$ pairs in a strong
magnetic field. The pairs screen the field $E_\parallel$ in the
pulsar magnetosphere everywhere except for some compact regions
called "gaps". These gaps are, in fact, "engines" responsible for
the non-thermal radiation of pulsars.

\par
Three kinds of gaps have been proposed in trying to explain the
non-thermal radiation of pulsars. The main difference between these
is in the site of gaps. A gap that forms near the magnetic pole of a
pulsar is called a polar gap (Ruderman \& Sutherland 1975). Besides,
an outer gap may form between the surface of null Goldreich-Julian
density and the light cylinder (Cheng, Ho, \& Ruderman 1986a,b). The
third possibility is a slot gap located in the space-charge-limited
flow along the last closed field line (Arons 1983; Muslimov \&
Harding 2004).

\par
In conventional models of polar gaps, it is assumed that created
$e^+e^-$ pairs are free (Ruderman \& Sutherland 1975; Daugherty \&
Harding 1982, 1996; Medin \& Lai 2010 and references therein).
However, this assumption is not valid if the magnetic field is
higher than $\sim 0.1\,B_0\simeq 4\times 10^{12}$ G, where
$B_0=m^2c^3/e\hbar=4.4\times 10^{13}$ G. The reason is that in such
a strong field, $\gamma$-rays emitted nearly along the curved field
lines, adiabatically convert into bound electron-positron pairs
(positronium) rather than decaying into free pairs (Shabad \& Usov
1985, 1986; Herold et al. 1985; Usov \& Melrose 1995, 1996; Baring
\& Harding 2001; Harding \& Lai 2006; Thompson 2008). The fact that
$e^+e^-$ pairs created in a magnetic field $B>0.1B_0$ are bound, may
be very important for the observational appearances of strongly
magnetized pulsars (Usov \& Melrose 1995). In particular, unlike
free pairs, such bound pairs do not screen the electric field
$E_\parallel$ near the pulsar, which requires a net charge density
to build up. As a result the pulsar luminosity is higher than it
would have been if the created pairs are free. For pulsars with
strong surface magnetic fields, $B_{_{\rm S}}> 0.1 B_{\rm cr}$, a
polar gap is modified by taking into account creation of bound pairs
as was suggested by Usov \& Melrose (1995, 1996). In this model, the
total power carried away by both relativistic particles and
radiation, from the polar gap into the magnetosphere, may increase
significantly (up to a few tens times) in comparison with the
conventional polar gap models, and it may be a substantial fraction
of the spin-down power. In this Letter, we demonstrate that the
total power of the modified polar gaps may be enough to explain the
observed luminosities of all known strongly magnetized, $\gamma$-ray
pulsars. Other observational consequences of the modified polar gap
model for these pulsars are also discussed.

\section{Radiative efficiency}

\par
In the modified polar gap model (Usov \& Melrose 1985, 1986), the
fraction of the spin-down power carried away by both relativistic
particles and radiation from the polar gap into the magnetosphere is
estimated as

$$\eta^b_\gamma= {L_{\rm p}^b\over\dot E_{\rm
rot}}\simeq {3\over2}\left({P\over P_1}\right)^{3/2}
\left[1-\left({P\over P_1}\right)^{3/2}\right]\,. \eqno(1)$$

\noindent where $P_1\simeq 0.5 ({B_{\rm p}/ 0.1\,B_0})^{2/3}$ s, and
$B_{\rm p}$ is the magnetic field strength at the poles. At
$P=2^{-2/3}P_1\simeq 0.63\,P_1$, the value of $\eta^b_\gamma$ is
maximum, $\eta^b_\gamma|_{\rm max}= {3/8}=0.375$. In this case the
polar gap luminosity $L_{\rm p}^b$ is comparable with the spin-down
power.

\par
The modified polar gap model is valid if both $B_{\rm p}>0.1B_0$ and
$P_2<P<P_1$ (Usov \& Melrose 1986), where
$$P_2\simeq 0.07 (T_{_S}/10^6\,{\rm
K})^{4/11}(B_{\rm p}/0.1B_{\rm cr})^{2/11}\,\,{\rm s}.\eqno(2)$$ For
$\gamma$-ray pulsars satisfied these conditions, measured and
derived properties are given in Table~1. Additionally, the
parameters of two $\gamma$-ray pulsars (J0633+1746 and J1057-5226)
are included in Table~1. In these two cases, the observed efficiency
of transformation of the spin-down power into the non-thermal
high-frequency emission is very high, $\eta_\gamma^{\rm
obs}=L_{\gamma +X}/\dot E_{\rm rot}\sim 1$, while the dipolar
B-field estimate is only slightly below the value required for
bound-pair formation. The surface magnetic field may be higher than
the dipolar estimate and satisfies the requirement of bound-pair
formation, $B_{\rm p}\geq 0.1B_0$, provided one invokes higher-order
multipolar components, or an off-centered dipole. To estimate the
possible value of $\eta^b_\gamma$ for these two pulsars we take
$B_{\rm p}=0.1B_0=4.4\times 10^{12}$ G (see the second lines in
Table 1).

\par
From Table~1 we can see that the predicted value of $\eta^b_\gamma$
is about or more than the minimum observed value of the radiative
efficiency, $\eta_\gamma^{\rm obs}$, for all strongly magnetized
$\gamma$-ray pulsars except of PSR J2021+4026. It is worth to note
that the $\gamma$-ray luminosities of pulsars given in Table~1 were
only roughly estimated by Abdo et. al. (2010) using the following
equation
$$ L_\gamma =4\pi d^2f_\Omega G\,,\eqno(3)$$
where $d$ is the distance of the pulsar, $G$ is the average energy
flux in $\gamma$-rays, and $f_\Omega$ is the flux correction factor
that takes into account the angular anisotropy of the $\gamma$-ray
emission of pulsars. The value of $f_\Omega$ is model-dependent and
varies from $\sim 1$ in the outer gap and slot gap models to $\sim
1/4\pi\simeq 0.08$ in the polar gap model (e.g., Thompson et al.
1994). Abdo et al. (2010) have used $f_\Omega=1$ throughout the
paper, and therefore, the $\gamma$-ray luminosities may be
overestimated at least a few times. Summarizing, the power carried
away by both relativistic particles and radiation from the modified
polar gaps into the magnetosphere may be sufficient to explain the
non-thermal luminosities of all known strongly magnetized
$\gamma$-ray pulsars, including PSR J2021+4026.

\par
For comparison, in conventional polar-gap models, where the  created
pairs are assumed to be free, the corresponding fraction of the
spin-down power going into both high-energy particles and radiation
may be estimated as (e.g., Usov \& Melrose 1995)

$$\eta_\gamma^f={L_{\rm p}^f\over\dot E_{\rm
rot}}\simeq 1.5\times10^{-3} \left({B_{\rm p}\over
0.1\,B_0}\right)^{-8/7} \left({P\over 0.1\,{\rm
s}}\right)^{15/7}.\eqno(4)$$ From Table 1 we can see that
$\eta^f_\gamma$ is more than an order of magnitude smaller than
$\eta_\gamma^{\rm obs}$ except for a few $\gamma$-ray pulsars. Even
taking into account the uncertainty in $\eta^{\rm obs}_\gamma$
because of the uncertainty of $f_\Omega$, we can see that for the
main part of the strongly magnetized $\gamma$-ray pulsars the
inferred high efficiency of conversion of rotational energy into
$\gamma$-ray radiation cannot be explained within the conventional
polar-gap models.

\par For six pulsars the observed luminosities in $\gamma$-rays are
not presented in Table~1 because their distances are unknown.
Equalizing $\eta^{\rm obs}_\gamma$ to $\eta^{\rm b}_\gamma$ for
these pulsars, upper limits on the distances may be estimates in the
frame of the modified polar gap model.

\section{Spectrum of $\gamma$-rays}

\par
The $\gamma$-ray spectra of pulsars were fitted by an exponentially
cutoff power-law model, $dN/dE\propto E^{-\alpha}\exp(-E/E_{\rm
cuoff})$, with the photon index $\alpha$ in the range 0.7-2.4 and
the cutoff energy $E_{\rm cuoff}$ at 0.1-7 GeV (Abdo et al. 2010,
Pilia et al. 2010). These spectra may be explained by curvature
radiation of high-energy electrons that escape from the polar caps
with the Lorentz factor $\Gamma\sim 10^7$ (Harding 1981, Daugherty
\& Harding 1982; Chiang \& Romani 1992). The value of $\Gamma\sim
10^7$ is more or less appropriate for primary electrons at the top
edge of the polar gap in the conventional polar gap models, in which
the density of primary particles is slightly smaller than the
Goldreich-Julian density $n_{\rm GJ}$ (Goldreich \& Julian 1969;
Ruderman \& Sutherland 1975; Arons 1981). To explain the
luminosities of $\gamma$-ray pulsars, it was suggested by Harding
(1981) that the density of the outflowing electrons with Lorentz
factor $\Gamma\sim 10^7$ is $\sim (10^2-10^3)n_{\rm GJ}$. Although
this suggestion is inconsistent with the conventional polar gap
models, it may be consistent with the modified polar gap model (Usov
\& Melrose 1995, 1996). Indeed, inside the modified polar gaps the
primary electrons are accelerated by the electric field
$E_\parallel\simeq E_\parallel^{\rm ion}$ to the Lorentz factor
$\Gamma_{\rm prim}\simeq (2-3)\times 10^8$ and generate very hard
$\gamma$-rays via curvature radiation, where $E_\parallel^{\rm
ion}\simeq (1-2)\times 10^{10}$ V/cm is the electric field at which
ield ionization of bound pairs becomes important. These
$\gamma$-rays are absorbed in the magnetic field and create
secondary electron-positron pairs with the mean Lorentz factor and
the density
$$\Gamma_{\rm s}\simeq 10^7 \left({P\over 0.1\,{\rm
s}}\right)^{1/4}\,{\rm and}\,\,n_{\rm s}\simeq 4\times 10^2
\left({P\over 0.1\,{\rm s}}\right)^{-3/4}n_{\rm GJ},\eqno(5)$$
respectively (Usov \& Melrose 1996). The parameters of the secondary
particles are well consistent with those suggested by Harding (1981)
to explain both the luminosities and spectra of $\gamma$-ray
pulsars.

\par
In the frame of the conventional polar gap models, it was argued
that photon splitting, $\gamma +B\rightarrow \gamma ' +\gamma ''
+B$, and magnetic absorption of photons, $\gamma +B\rightarrow
e^++e^-+B$, have to reduce the cutoff energy $E_{\rm cutoff}$ for
strongly magnetized $\gamma$-ray pulsars, $B_{\rm p}\gtrsim
10^{13}$~G (Harding et al. 1997; Baring \& Harding 2001; Baring
2004). This is consistent with the observational data on PSR
B1509-58 that has one of the highest magnetic fields, $B_{\rm
p}\simeq 3.1\times 10^{13}$ G, and the softest spectrum with a
cutoff at $E_{\rm cutoff}=81\pm 20$ MeV. The second softest
$\gamma$-ray spectrum with $E_{\rm cutoff}=0.7\pm 0.5$ GeV is that
of PSR B0656+14, recently observed by {\it Fermi}. This pulsar has a
rather strong magnetic field, $B_{\rm p}\simeq 0.93\times 10^{13}$
G, and its soft $\gamma$-ray spectrum may be explained in the same
way. At present, however, there are a handful of $\gamma$-ray
pulsars with surface magnetic fields higher than that at the surface
of PSR B0656+14, while their $\gamma$-ray emission is present to at
least a few GeV. One such pulsar is LAT PSR J0007+7303, in the
supernova remnant CTA 1. This pulsar has one of the hardest
$\gamma$-ray spectra with $E_{\rm cutoff}=4.6\pm 0.4$ GeV, while the
magnetic field at its poles is $B_{\rm p}\simeq 2.2\times 10^{13}$
G, which is only slightly smaller than the $B_{\rm p}$-value for PSR
B1509-58. Among other pulsars with hard ($E_{\rm cutoff} > $ a few
GeV) $\gamma$-ray spectra and strong ($B_{\rm p}> 0.1 B_0$) magnetic
fields are PSR J0631+1036, J0633+1746, J1709-4429, and J2021+4026.
Since one expects strong absorption of hard $\gamma$-rays in the
vicinity of these pulsars (Harding et al. 1997, Baring 2004), it is
concluded that (1) a polar gap model is inconsistent with the
available data on the strongly magnetized $\gamma$-ray pulsars with
hard $\gamma$-ray spectra, and (2) the $\gamma$-ray emission of
these pulsars arises largely in the outer magnetosphere (e.g., Abdo
et al. 2010; Michelson, Atwood, \& Ritz 2010).

However, these conclusions may be premature at least for some
strongly magnetized $\gamma$-pulsars. Indeed, in the modified polar
gap model $\gamma$-rays generated near the surface of a strongly
magnetized pulsar adiabatically convert into bound $e^+e^-$ pairs.
If the photoionization of the bound pairs is small, these pairs
mostly flow away from the pulsar vicinity and may annihilate at
large distances generating hard $\gamma$-rays that escape from the
pulsar magnetosphere. In other words, bound pairs may be an
intermediate agent that transfers $\gamma$-rays from the pulsar
vicinity to the outer magnetosphere where their absorption is
negligible.

\par
The surface temperature of PSR B1509-58 may be as high as $\sim
2\times 10^6$ K. Besides, this pulsar is a powerful ($L_X\simeq
10^{35}$ erg/s) source of soft X-rays with a non-thermal (power-law)
spectrum. The soft X-ray and $\gamma$-ray pulses are phase-aligned,
indicating that they are generated in the same region in the pulsar
magnetosphere (near the magnetic poles in the polar gap model). Most
probably, the bound pairs are mostly ionized by both thermal and
non-thermal X-rays before leaving the vicinity of PSR B1509-58. In
this case, the formation of bound pairs does not affect the
conclusion of Harding et al. (1997) that in the polar gap model the
$\gamma$-ray spectrum of PSR B1509-58 is expected to be very soft.
As to PSR J0007+7303, there is an upper limit of $T_{\rm S}<
6.6\times 10^5$ K (Halpern et al. 2004). Recently, Caraveo et al.
(2010) decreased the upper limit on the surface temperature of PSR
J0007+7303: $T_{\rm S}< 5.3\times 10^5$ K, that is one of the most
constraining data points on cooling models of neutron stars. This
makes PSR J0007+7303 by far the coldest neutron star for its age
interval, suggesting the necessity of enhanced neutrino emission for
this rather young $\gamma$-ray pulsar (Page, Geppert, \& Weber 2006;
Caraveo et al. 2010). The surface temperature of PSR J0007+7303 in
the cooling model of neutron stars with enhanced neutrino emission
may be as low as $(1-2)\times 10^5$ K (Page et al. 2006). For this
temperature, the bound pairs may leave the pulsar vicinity and
annihilate far from the pulsar into hard$\gamma$-rays,
$\varepsilon_\gamma\gtrsim$ a few GeV.

\section{Discussion}

\par In this paper, we have demonstrated that the observed
non-thermal luminosities of strongly magnetized pulsars may be
explained in the polar gap model modified by taking into account
the process of adiabatical conversion of $\gamma$-rays into bound
electron-positron pairs in a strong magnetic field, $B\gtrsim
4\times 10^{12}$ G. The high-energy spectra of $\gamma$-ray
pulsars may be also explained by the modified polar gap model.
However, all available polar gap models have difficulty in
reconciling the observed $\gamma$-ray pulses with the known radio
pulsar geometry. At present, it is not clear possible or not to
overcome this difficulty in the modified polar gap model at least
for some strongly magnetized pulsars.

\begin{table}
\caption{Properties of $\gamma$-ray pulsars}
\begin{center}

$$\vbox{
\tabskip=0pt \halign{ \hfil#\hfil& \hfil$\;#\;$\hfil&
\hfil$\;#\;$\hfil& \hfil$\;#\;$\hfil& \hfil$\;#\;$\hfil&
\hfil$\;#\;$\hfil& \hfil$\;#\;$\hfil& \hfil$\;#\;$\hfil&
\hfil$\;#\;$\hfil \cr \noalign{\hrule\vskip5pt} PSR   & P  & B_{\rm
p}
   &  {\rm Age }  &   \dot E_{\rm rot}    &
    L_{\gamma +X}  &  \eta^{\rm obs}_\gamma& \eta^f_\gamma & \eta^b_\gamma
\cr \noalign{\vskip3pt}
&({\rm ms})&(10^{12}{\rm G})&({\rm kyr})& (10^{34}{\rm erg\,
s^{-1}})& (10^{34}{\rm erg\, s^{-1}}) & (10^{-2})
&(10^{-2})&(10^{-2}) \cr \noalign{\vskip5pt\hrule\vskip5pt}
J0007+7303 &  316  &  22   &   14    &  45.2  & 8.9\pm 3.8 &  20\pm
8 & 0.28 & 14 \cr
J0248+6021 &  217   &   7  &   63     & 21    & 1.5-30 & 7-140 &
0.46 &  22  \cr
J0357+32  &   444  &  4.7  &   590  &   0.5    &  -
 & - & 3.4 & 25  \cr
J0631+1036  & 288  &  11 &   44  & 17.3   &  0.2-4.8  & 1-27 & 0.51
& 22 \cr
J0633+0632  & 297  &  10  &  59  & 11.9 & - & -  & 0.6 & 24\cr
J0633+1746  & 237  &  3.3  & 340  & 3.3 & 2.5^{+2.4}_{-1.2} &
78^{+74}_{-38} & 1.1 &  -\cr
& ...& (4.4) & ... & ... & ... &... & 1 & 33\cr
J0659+1414  & 385  &  9.3  & 110  & 3.8 & 0.031\pm 0.008 &  1\pm 0.2
& 1.1 &  33  \cr
J1048-5832 &  124  &  7 & 20 &  201 & 15\pm 9  & 8\pm 5 & 0.14&
 11 \cr
J1057-5226 &  197  &  2.2 & 540 &  3.0 & 1.7\pm 0.9  & 56\pm 31 &
1.2&
 - \cr
&  ... & (4.4) & ... & ... & ... & ... & 0.64 & 28\cr
J1124-5916 &  135  &  20 & 3 &  1190 & 10^{+3.4}_{-5.4}  &
1^{+0.3}_{-0.4} & 0.05 & 4.2 \cr
J1418-6058 &  111  &  8.8 & 10 &  498 & 11-70  & 2-14 & 0.08 &
 7.4 \cr
B1509-58 &   150  &  31  & 1.7 & 1800  & 39\pm 14   & 2.2\pm 0.8
&0.04 & 3.4 \cr
J1709-4429 &  102  & 6.3 & 18 & 341 & 29-190 & 9-57  & 0.1 & 9 \cr
J1732-31 &   197  &  4.6 & 120 & 13.6  &  -  & -  & 0.61 & 27 \cr
J1741-2054 &  414  &  5.4 & 390 & 0.9  &  0.22\pm 0.13  & 24\pm 14 &
2.5 &  36  \cr
J1809-2332 &  147  &  4.6 & 68 &  43 & 2-28  & 4.6-66 & 0.33 &
 19 \cr
J1826-1256 &  110  &  7.4 & 14 &  358 & -  & - & 0.1 &
 8.6 \cr
J1958+2846  &  290  & 16 & 21  & 35.8  &  -  & -  & 0.34 & 16 \cr
J2021+3651&   104  &  6.4 & 17 & 338 &  1-75  & 0.3-22 & 0.1 & 9.2
\cr
J2021+4026 &   265  &  7.7 & 77 & 11.6  &  26\pm 15  & 220\pm 130 &
0.64 & 26  \cr
J2238+59 &   163  &  8.1 & 26 & 90.3  &  -  & - & 0.21  & 14 \cr
\noalign{\vskip5pt\hrule} } }$$

\end{center}
\end{table}

\end{document}